\documentclass[12pt]{article}\usepackage[]{graphicx}\usepackage[]{color}
\usepackage[top=1in,left=1in, right = 1in, footskip=1in]{geometry}

\title{A note on observation processes in epidemic models}
\author{Sang Woo Park and Benjamin M. Bolker}

\usepackage{tabularx}

\usepackage{amsmath}
\usepackage{natbib}
\usepackage{hyperref}
\date{\today}

\usepackage{bm}

\usepackage{afterpage}
\usepackage{pdflscape}

\begin{document}

\begin{flushleft}{
	\Large
	\textbf\newline{
		A note on observation processes in epidemic models
	}
}
\newline
\\
Sang Woo Park\textsuperscript{1,2,*}
Benjamin M. Bolker\textsuperscript{2,3,4}
\\

\bigskip
\textbf{1} Department of Ecology and Evolutionary Biology, Princeton University, Princeton, New Jersey, USA
\\
\textbf{2} Department of Mathematics and Statistics, McMaster University, Hamilton, Ontario, Canada
\\
\textbf{3} Department of Biology, McMaster University, Hamilton, Ontario, Canada
\\
\textbf{4} Michael G. DeGroote Institute for Infectious Disease Research, McMaster University, Hamilton, Ontario, Canada
\\
\bigskip

*swp2@princeton.edu
\end{flushleft}

\section*{Abstract}

Many disease models focus on characterizing
the underlying transmission mechanism but make simple, possibly naive assumptions about
how infections are reported. In this note, we use a simple deterministic
Susceptible-Infected-Removed (SIR) model to compare two common assumptions 
about disease incidence reports: individuals can report their infection as soon as
they become infected or as soon as they recover. We show that
incorrect assumptions about the underlying observation processes can bias
estimates of the basic reproduction number and lead to overly
narrow confidence intervals.

\section{Introduction}

Mechanistic analyses of epidemic time series allow us to make inference about the underlying 
transmission mechanisms, estimate biologically relevant parameters, and 
predict the course of an outbreak \citep{breto2009time}. In order to make 
precise and accurate inferences, disease modelers have sought to build
more realistic process models. For example, the time series of reported
measles cases from London in the prevaccination era has been analyzed many times, using
variety of models accounting for time-varying transmission rates \citep{fine1982measles}, 
realistic age structure \citep{schenzle1984age},
metapopulation structure \citep{xia2004measles}, continuous-time infection processes 
\citep{cauchemez2008likelihood}, and extra-demographic variability \citep{he2009plug}. 

Despite the amount of effort put into developing better process models, 
disease modelers often neglect details of the observation processes associated with
new disease case reports (often referred to as incidence time series).
Many disease models effectively assume that new cases are reported
instantaneously when an individual is infected (e.g., \cite{martinez2016differential, 
kennedy2018modeling, pons2018serotype}) or when an individual becomes symptomatic
(e.g., \cite{bhadra2011malaria, king2015avoidable}); 
some models (e.g., \cite{breto2009time, he2009plug, lin2016seasonality})
assume that infections are counted upon recovery (because diagnosed cases
are controlled and are effectively no longer infectious).

We emphasize that incidence (i.e., the number of \emph{newly} infected individuals) is different from prevalence (i.e., the number of \emph{currently} infected individuals) \citep{bjornstad2018epidemics}.
The dynamics of incidence depend on the reporting time step (because the sum of true incidence is equal to the final size of an epidemic), whereas those of prevalence do not.
We expect the dynamics of incidence and prevalence to be similar only when the reporting time step is similar to the disease generation time \citep{fine1982measles}.
While they are uncommon, some models do not make a clear distinction between prevalence and 
incidence \citep{capistran2009parameter, hooker2010parameterizing, yang2013stability, gonzalez2014fractional}.

Here, we use a simple Susceptible-Infected-Removed (SIR) model to study how
assumptions about the underlying observation processes affect parameter estimates
of the SIR model. We show that making incorrect assumptions about the timing of 
incidence reports can lead to biased parameter estimates and overly narrow 
confidence intervals.

\section{Methods}

The Susceptible-Infected-Removed (SIR) model describes how a disease spreads in a
homogeneous population:
\begin{equation}
\begin{aligned}
\frac{dS}{dt} &= - \beta S \frac{I}{N}\\
\frac{dI}{dt} &= \beta S \frac{I}{N} - \gamma I\\
\frac{dR}{dt} &= \gamma I,
\end{aligned}
\end{equation}
where $\beta$ is the contact rate per unit time, $\gamma$ is the recovery rate per unit time, 
and $N = S + I + R$ is the total population size. 
We define \emph{true} incidence at time $t$ as the number of newly infected
individuals that are infected between time $t- \Delta t$ and time $t$, where $\Delta t$ is
the reporting time step. We expect infected cases to be reported some time after infection;
the number of reported cases during a time period defines the \emph{observed} incidence

For brevity, we consider two extreme cases: individuals instantaneously report
their infection when they become infected or when they recover. The observed incidence 
measured upon infection, $i_1(t)$, can be defined by the integral:
\begin{equation}
i_1(t) = \int_{t - \Delta t}^{t} \beta S \frac{I}{N} dt.
\end{equation}
Equivalently, we can keep track of cumulative incidence, $C$, by adding a 
new state variable described by $dC/dt = \beta S I/N$ and taking the difference between 
the two consecutive reporting periods: $i_1(t) = C(t) - C(t-\Delta t)$. Likewise, 
the observed incidence measured upon recovery, $i_2(t)$, can be defined by the integral:
\begin{equation}
i_2(t) = \int_{t-\Delta t}^{t} \gamma I dt,
\end{equation}
or by the consecutive difference in the cumulative number of recovered cases:
$i_2(t) = R(t) - R(t - \Delta t)$.
Finally, we model observation error using a negative binomial distribution with a
mean of either $\rho i_1(t)$ or $\rho i_2(t)$, where $\rho$ is the reporting rate, and
an over-dispersion parameter $\theta$. For convenience, we will refer to these two
negative binomial models as infection model and recovery model hereafter; 
similarly, we will refer to epidemic time series generated from these
negative binomial models with two different means ($\rho i_1(t)$ and $\rho i_2(t)$)
as infection time series and recovery time series.

In this study, we focus on estimating 5 parameters: the basic reproductive
number $\mathcal R_0 = \beta/\gamma$, mean infectious period $1/\gamma$, 
reporting rate $\rho$, the overdispersion parameter $\theta$, and the initial
proportion of the infected individuals $i_0$. The initial proportion of
susceptible individuals is assumed to be $1 - i_0$. The total population
size $N$ is assumed to be known.

\section{Results}

\begin{figure}[!t]
\includegraphics[width=\textwidth]{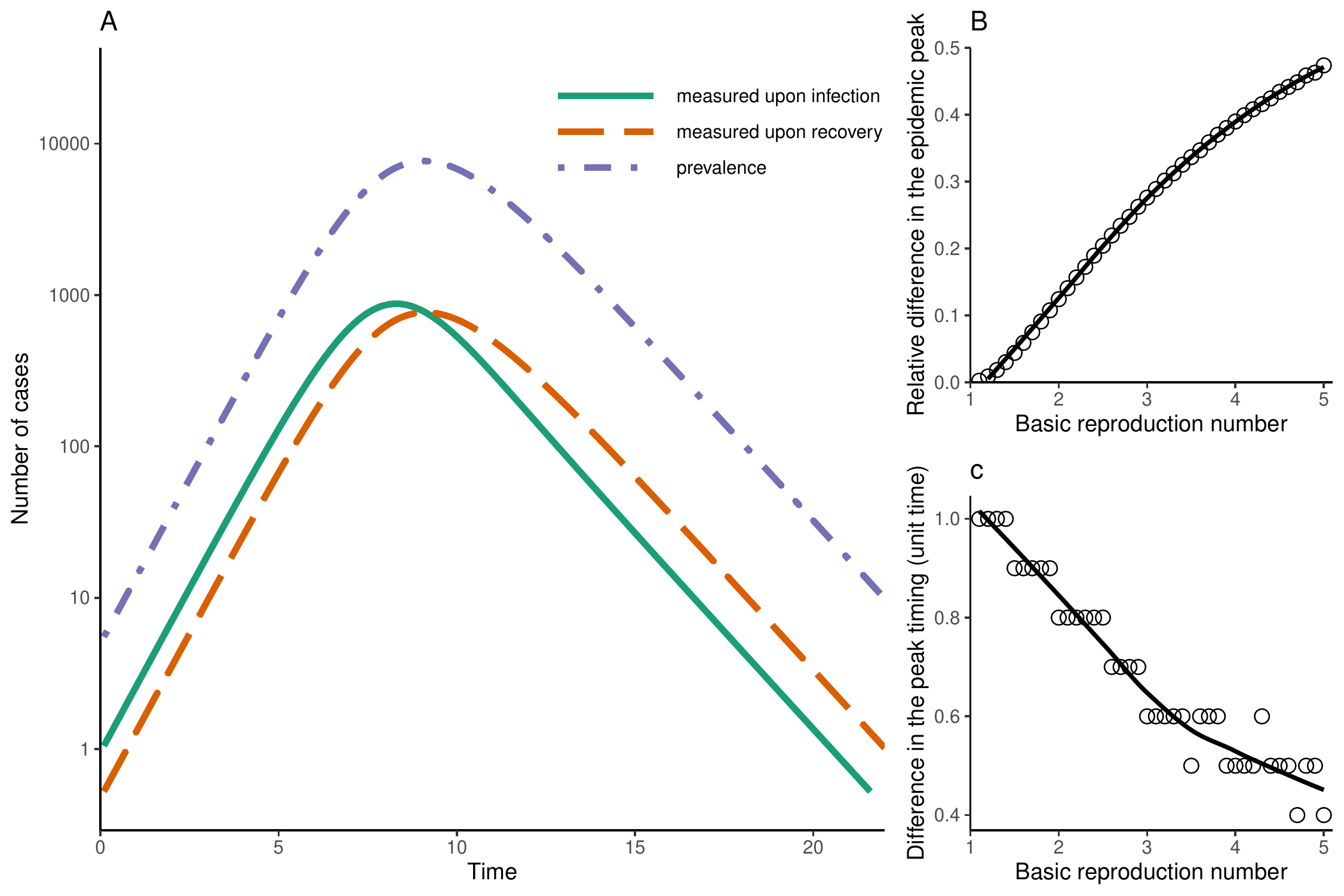}
\caption{
\textbf{A comparison of incidence measured at two different time points in infection.}
(A) 
A deterministic simulation of the SIR model using the following parameters: 
$\mathcal R_0 = 2$, $1/\gamma = 1$ time units, $N = 1 \times 10^5$, $i0 = 1 \times 10^{-4}$,
and $\Delta t = 0.1$ time units.
(B-C) Effects of $\mathcal R_0$ on the relative difference in the size of the epidemic peak ($1 - [\max i_2(t)]/[\max i_1(t)]$) and the difference in the peak timing ($\hat{t}_2 - \hat{t}_1$ where $i_k(\hat{t}_k) = \max i_k(t)$ for $k = 1, 2$) due to delays in reporting time.
Open circles: simulation results of the deterministic SIR model.
Solid lines: locally estimated scatterplot smoothing (LOESS) curves fitted to simulation results.
Remaining parameters are held constant ($1/\gamma = 1$ time units, $N = 1 \times 10^5$, $i0 = 1 \times 10^{-4}$, and $\Delta t = 0.1$ time units) throughout simulations.
}
\end{figure}

Figure 1A compares the deterministic dynamics of two incidence curves,
$i_1(t)$ and $i_2(t)$, and a prevalence curve $I(t)$ for $\mathcal R_0 = 2$. 
A lag in reporting time delays
the timing of the observed epidemic peak and reduces the size of that peak.
As $\mathcal R_0$ increases, the difference in the size of the peaks increases (Figure 1B)
but the difference in the timing of the peaks decreases (Figure 1C).
For example, when $\mathcal R_0 = 5$, such delay in the reporting of new cases can reduce the size of the observed epidemic peak by almost 50\%.
Note that $i_2(t)$ essentially assumes that the amount of time between when an individual is infected and when an individual reports the infection is exponentially distributed; a fixed delay in reporting time will not change the shape of an epidemic curve.

For small values of $\mathcal R_0$, 
these differences in the reporting time have little effect on the overall shape 
of the epidemic curve. In the presence of observation and process error, we
do not expect to be able to distinguish between the two reporting processes
based on the time series alone (and we rarely know \emph{a priori} the delay distribution between when an individual is infected and when that case is reported).
Therefore, one might naively expect assumptions about the timing of case reporting to have negligible effect on inference.

In order to understand how assumptions about the timing of case reporting affect 
parameter estimates of the SIR model, we simulate epidemic time series
(infection time series and recovery time series) 
100 times with $\mathcal R_0 = 2$ and fit both infection and recovery models to each 
time series. We compare the estimates of the basic reproductive number $\mathcal R_0$,
and the coverage of our confidence intervals, defined as the proportion of confidence intervals that
contain the true value (95\% confidence interval is expected to contain the true value
95\% of the time by definition). Figure 2 summarizes the results.

When we try to estimate all 5 parameters, fitting the recovery model to
infection time series underestimates the basic reproduction number and 
gives a slightly low coverage (Figure 2A). 
Fitting the infection model to recovery
time series slightly overestimates the basic reproduction number but gives 
good coverage. 
We expect fitting incorrect models to give more biased estimates when $\mathcal R_0$ is higher because the differences in the size of observed epidemic peaks become greater.
Fitting the correct model gives unbiased estimates and good coverage.

\begin{figure}
\includegraphics[width=\textwidth]{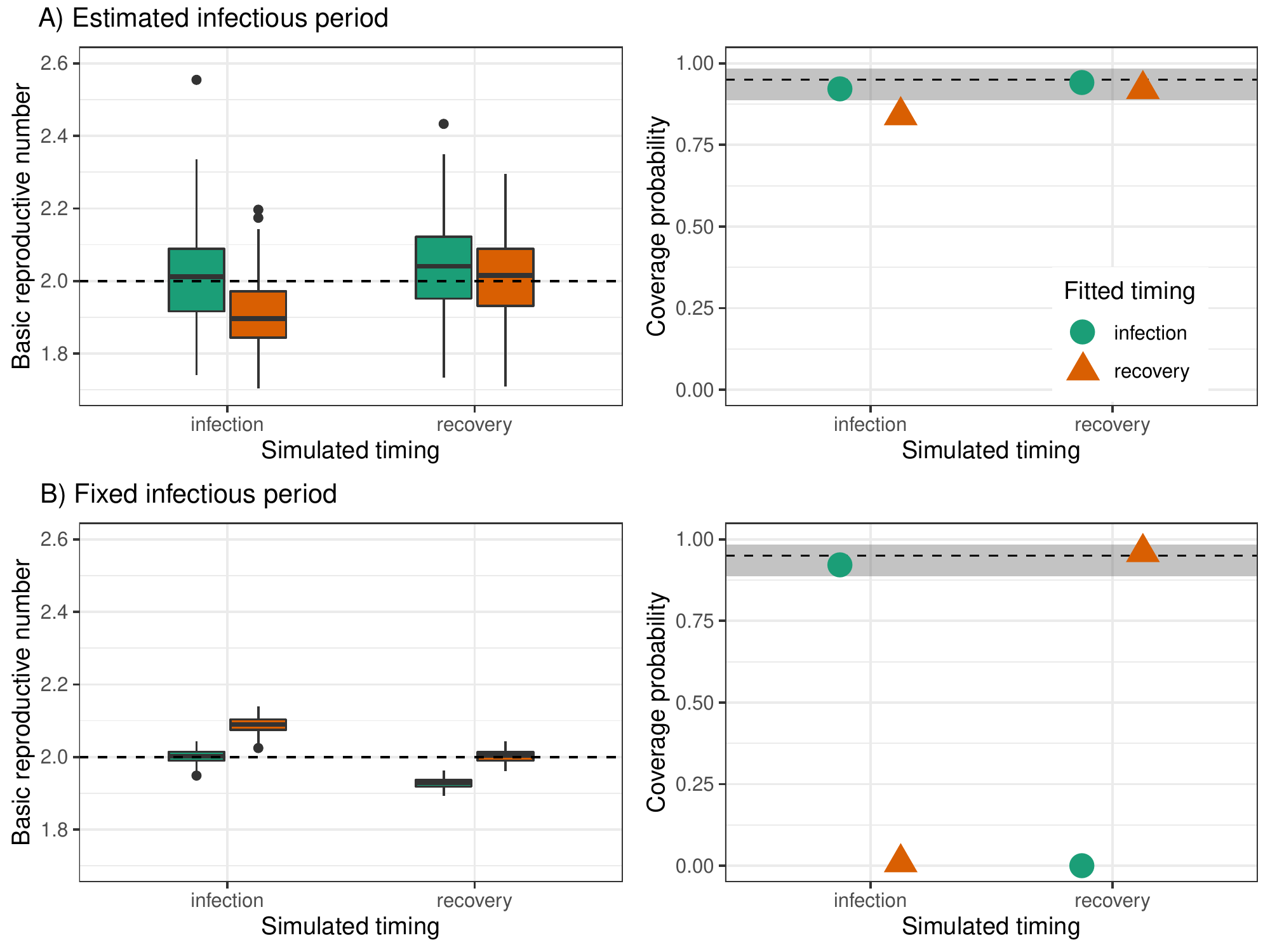}
\caption{
\textbf{A comparison of incidence measured at two different time points in infection.}
We simulate infection time series and recovery time series 100 times using the 
following parameters:  
$\mathcal R_0 = 2$, $1/\gamma = 1$ time units, $N = 1 \times 10^5$, $i0 = 1 \times 10^{-4}$,
$\rho = 0.5$, $\theta = 10$, and $\Delta t = 0.1$ time units.
For each simulation, we fit the infection model and the recovery model by
(A) estimating the mean infectious period and (B) assuming
that the mean infectious period is known.
Coverage probabilities represent the proportion of confidence intervals
that contain the true value of the basic reproductive number ($\mathcal R_0$).
}
\end{figure}

Disease modelers often assume that the mean infectious period of a disease
is known and focus on estimating the basic reproduction number (e.g.,
\cite{hooker2010parameterizing, lin2016seasonality, pons2018serotype}). 
When we assume that the true value of the mean infectious period is known
and try to estimate the remaining 4 parameters of the SIR model (Figure 2B), fitting
incorrect models results in a clearer bias (in opposite directions)
and a much lower coverage (cf. \cite{elderd2006uncertainty}).

Differences in the direction of the bias can be explained by the estimates of the
exponential growth rate ($r = \beta - \gamma$) and the mean infectious period ($1/\gamma$).
In general, we expect delays in observation processes to make the observed epidemic 
time series last longer and have smaller peaks (Figure 1).
When we fit the recovery model to an infection time series, the model overestimates
the initial growth rate in order to match the bigger (and faster) epidemic peak of the 
infection time series. When the mean infectious period is fixed, higher growth rate
translates to higher basic reproduction number, as Figure 2B shows. When we allow
the mean infectious period to vary, the model still overestimates the growth rate but
also underestimates the mean infectious period (high $\gamma$), 
which decreases the overall estimate of the basic reproduction number.
Similarly, fitting the infection model to a recovery time series underestimates
the growth rate to match the smaller (and slower) epidemic peak; this 
underestimates the basic reproduction number when the mean infectious period is fixed. 
When we allow the mean infectious period to vary, we overestimate the mean infectious
period, which in turn increases the estimate of the basic reproduction number.

\section{Discussion}

Mathematical modeling of infectious disease outbreaks helps us understand how disease spreads in a population; 
however, epidemic models often make simple assumptions about how cases are reported. 
We used a deterministic SIR model to show that delays in case reports affect the observed shape of an epidemic curve and the estimates of the basic reproduction number.
Even when the basic reproduction number is small (e.g., $\mathcal R_0 = 2$), fitting incorrect models can introduce small bias and give narrow confidence intervals.

We compared two scenarios in which newly infected cases are reported 
instantaneously (1) when individuals become infected or
(2) when they recover. Although neither of these
assumptions is realistic, many epidemic models still rely on these
assumptions (see Introduction). 
More realistic models may distinguish reported and unreported
(i.e., identified and unidentified) cases by adding
new state variables \citep{browne2015modeling,webb2015model} 
or by modeling an explicit delay distribution in reporting time 
\citep{harris1990reporting, ferguson2001foot, goldstein2009reconstructing,
ster2009epidemiological, birrell2011bayesian, funk2018real}.

We considered observation processes associated with incidence 
reporting; however, we expect observation processes to be just as 
(if not more) important in analyzing mortality data. Many disease modelers have
tried to infer underlying transmission mechanisms from historical mortality 
data but assumed that individual deaths are recorded as soon as individuals die 
\citep{he2013inferring, didelot2017model, dean2018human}; this includes the classic work by
\cite{kermack1927contribution} who approximated the reported \emph{number} of
deaths per week from plague with instantaneous death \emph{rates} ($dR/dt$).
These frameworks do not account for the possibility 
that a delay in reporting of deaths can
change the shape of an epidemic curve -- delays in case reports can decrease
the size of the observed epidemic peak and delay the observed timing of the peak (Figure 1). 

Here, we assumed that the underlying transmission process is deterministic; this 
assumes that all error can be explained by observation errors alone. 
We chose to study a deterministic model SIR for computational efficiency; 
we do not recommend using deterministic models for real outbreak analyses.
Ignoring process errors (i.e., stochasticity in the transmission process) can lead to
overly confident results \citep{king2015avoidable}. Using stochastic models may
give better coverage probabilities even when wrong observation models are used.
Nonetheless, misspecifying the observation model may still affect conclusions from 
stochastic models and introduce bias.

Our study shows that a seemingly negligible change in the assumptions of an epidemic
model can affect the inference of infectious disease transmission.
While the amount of bias introduced from the misspecification of the observation
model can be fairly small (e.g., less than 5\% in our examples), any modelers
trying to make serious forecasts should be aware of potential biases.
We caution disease modelers to carefully consider the implications of their model assumptions when developing epidemic models.

\section*{Acknowledgements}

We thank David Earn for providing helpful comments on the manuscript. BMB is supported by an NSERC Discovery grant.

\bibliography{observation}

\begin{thebibliography}{}

\bibitem[\protect\citeauthoryear{Bhadra, Ionides, Laneri, Pascual, Bouma, and
  Dhiman}{Bhadra et~al.}{2011}]{bhadra2011malaria}
Bhadra, A., E.~L. Ionides, K.~Laneri, M.~Pascual, M.~Bouma, and R.~C. Dhiman
  (2011).
\newblock {Malaria in Northwest India: Data analysis via partially observed
  stochastic differential equation models driven by L{\'e}vy noise}.
\newblock {\em Journal of the American Statistical Association\/}~{\em
  106\/}(494), 440--451.

\bibitem[\protect\citeauthoryear{Birrell, Ketsetzis, Gay, Cooper, Presanis,
  Harris, Charlett, Zhang, White, Pebody, et~al.}{Birrell
  et~al.}{2011}]{birrell2011bayesian}
Birrell, P.~J., G.~Ketsetzis, N.~J. Gay, B.~S. Cooper, A.~M. Presanis, R.~J.
  Harris, A.~Charlett, X.-S. Zhang, P.~J. White, R.~G. Pebody, et~al. (2011).
\newblock {Bayesian modeling to unmask and predict influenza A/H1N1pdm dynamics
  in London}.
\newblock {\em Proceedings of the National Academy of Sciences\/}~{\em
  108\/}(45), 18238--18243.

\bibitem[\protect\citeauthoryear{Bj{\o}rnstad}{Bj{\o}rnstad}{2018}]{bjornstad2018epidemics}
Bj{\o}rnstad, O.~N. (2018).
\newblock {\em Epidemics: models and data using {R}}.
\newblock Springer.

\bibitem[\protect\citeauthoryear{Bret{\'o}, He, Ionides, King,
  et~al.}{Bret{\'o} et~al.}{2009}]{breto2009time}
Bret{\'o}, C., D.~He, E.~L. Ionides, A.~A. King, et~al. (2009).
\newblock Time series analysis via mechanistic models.
\newblock {\em The Annals of Applied Statistics\/}~{\em 3\/}(1), 319--348.

\bibitem[\protect\citeauthoryear{Browne, Gulbudak, and Webb}{Browne
  et~al.}{2015}]{browne2015modeling}
Browne, C., H.~Gulbudak, and G.~Webb (2015).
\newblock Modeling contact tracing in outbreaks with application to {Ebola}.
\newblock {\em Journal of theoretical biology\/}~{\em 384}, 33--49.

\bibitem[\protect\citeauthoryear{Capistr{\'a}n, Moreles, and
  Lara}{Capistr{\'a}n et~al.}{2009}]{capistran2009parameter}
Capistr{\'a}n, M.~A., M.~A. Moreles, and B.~Lara (2009).
\newblock Parameter estimation of some epidemic models. {The} case of recurrent
  epidemics caused by respiratory syncytial virus.
\newblock {\em Bulletin of mathematical biology\/}~{\em 71\/}(8), 1890.

\bibitem[\protect\citeauthoryear{Cauchemez and Ferguson}{Cauchemez and
  Ferguson}{2008}]{cauchemez2008likelihood}
Cauchemez, S. and N.~M. Ferguson (2008).
\newblock Likelihood-based estimation of continuous-time epidemic models from
  time-series data: application to measles transmission in {London}.
\newblock {\em Journal of the Royal Society Interface\/}~{\em 5\/}(25),
  885--897.

\bibitem[\protect\citeauthoryear{Dean, Krauer, Wall{\o}e, Lingj{\ae}rde,
  Bramanti, Stenseth, and Schmid}{Dean et~al.}{2018}]{dean2018human}
Dean, K.~R., F.~Krauer, L.~Wall{\o}e, O.~C. Lingj{\ae}rde, B.~Bramanti, N.~C.
  Stenseth, and B.~V. Schmid (2018).
\newblock {Human ectoparasites and the spread of plague in Europe during the
  Second Pandemic}.
\newblock {\em Proceedings of the National Academy of Sciences\/}~{\em
  115\/}(6), 1304--1309.

\bibitem[\protect\citeauthoryear{Didelot, Whittles, and Hall}{Didelot
  et~al.}{2017}]{didelot2017model}
Didelot, X., L.~K. Whittles, and I.~Hall (2017).
\newblock Model-based analysis of an outbreak of bubonic plague in {Cairo} in
  1801.
\newblock {\em Journal of The Royal Society Interface\/}~{\em 14\/}(131),
  20170160.

\bibitem[\protect\citeauthoryear{Elderd, Dukic, and Dwyer}{Elderd
  et~al.}{2006}]{elderd2006uncertainty}
Elderd, B.~D., V.~M. Dukic, and G.~Dwyer (2006).
\newblock Uncertainty in predictions of disease spread and public health
  responses to bioterrorism and emerging diseases.
\newblock {\em Proceedings of the National Academy of Sciences\/}~{\em
  103\/}(42), 15693--15697.

\bibitem[\protect\citeauthoryear{Ferguson, Donnelly, and Anderson}{Ferguson
  et~al.}{2001}]{ferguson2001foot}
Ferguson, N.~M., C.~A. Donnelly, and R.~M. Anderson (2001).
\newblock {The foot-and-mouth epidemic in Great Britain: pattern of spread and
  impact of interventions}.
\newblock {\em Science\/}~{\em 292\/}(5519), 1155--1160.

\bibitem[\protect\citeauthoryear{Fine and Clarkson}{Fine and
  Clarkson}{1982}]{fine1982measles}
Fine, P.~E. and J.~A. Clarkson (1982).
\newblock {Measles in England and Wales—I: an analysis of factors underlying
  seasonal patterns}.
\newblock {\em International journal of epidemiology\/}~{\em 11\/}(1), 5--14.

\bibitem[\protect\citeauthoryear{Funk, Camacho, Kucharski, Eggo, and
  Edmunds}{Funk et~al.}{2018}]{funk2018real}
Funk, S., A.~Camacho, A.~J. Kucharski, R.~M. Eggo, and W.~J. Edmunds (2018).
\newblock Real-time forecasting of infectious disease dynamics with a
  stochastic semi-mechanistic model.
\newblock {\em Epidemics\/}~{\em 22}, 56--61.

\bibitem[\protect\citeauthoryear{Goldstein, Dushoff, Ma, Plotkin, Earn, and
  Lipsitch}{Goldstein et~al.}{2009}]{goldstein2009reconstructing}
Goldstein, E., J.~Dushoff, J.~Ma, J.~B. Plotkin, D.~J. Earn, and M.~Lipsitch
  (2009).
\newblock Reconstructing influenza incidence by deconvolution of daily
  mortality time series.
\newblock {\em Proceedings of the National Academy of Sciences\/}~{\em
  106\/}(51), 21825--21829.

\bibitem[\protect\citeauthoryear{Gonz{\'a}lez-Parra, Arenas, and
  Chen-Charpentier}{Gonz{\'a}lez-Parra et~al.}{2014}]{gonzalez2014fractional}
Gonz{\'a}lez-Parra, G., A.~J. Arenas, and B.~M. Chen-Charpentier (2014).
\newblock A fractional order epidemic model for the simulation of outbreaks of
  influenza {A (H1N1)}.
\newblock {\em Mathematical methods in the Applied Sciences\/}~{\em 37\/}(15),
  2218--2226.

\bibitem[\protect\citeauthoryear{Harris}{Harris}{1990}]{harris1990reporting}
Harris, J.~E. (1990).
\newblock Reporting delays and the incidence of {AIDS}.
\newblock {\em Journal of the American Statistical Association\/}~{\em
  85\/}(412), 915--924.

\bibitem[\protect\citeauthoryear{He, Dushoff, Day, Ma, and Earn}{He
  et~al.}{2013}]{he2013inferring}
He, D., J.~Dushoff, T.~Day, J.~Ma, and D.~J. Earn (2013).
\newblock {Inferring the causes of the three waves of the 1918 influenza
  pandemic in England and Wales}.
\newblock {\em Proceedings of the Royal Society B: Biological Sciences\/}~{\em
  280\/}(1766), 20131345.

\bibitem[\protect\citeauthoryear{He, Ionides, and King}{He
  et~al.}{2009}]{he2009plug}
He, D., E.~L. Ionides, and A.~A. King (2009).
\newblock Plug-and-play inference for disease dynamics: measles in large and
  small populations as a case study.
\newblock {\em Journal of the Royal Society Interface\/}~{\em 7\/}(43),
  271--283.

\bibitem[\protect\citeauthoryear{Hooker, Ellner, Roditi, and Earn}{Hooker
  et~al.}{2010}]{hooker2010parameterizing}
Hooker, G., S.~P. Ellner, L.~D.~V. Roditi, and D.~J. Earn (2010).
\newblock Parameterizing state--space models for infectious disease dynamics by
  generalized profiling: measles in {Ontario}.
\newblock {\em Journal of The Royal Society Interface\/}~{\em 8\/}(60),
  961--974.

\bibitem[\protect\citeauthoryear{Kennedy, Dunn, and Read}{Kennedy
  et~al.}{2018}]{kennedy2018modeling}
Kennedy, D.~A., P.~A. Dunn, and A.~F. Read (2018).
\newblock {Modeling Marek's disease virus transmission: A framework for
  evaluating the impact of farming practices and evolution}.
\newblock {\em Epidemics\/}~{\em 23}, 85--95.

\bibitem[\protect\citeauthoryear{Kermack and McKendrick}{Kermack and
  McKendrick}{1927}]{kermack1927contribution}
Kermack, W.~O. and A.~G. McKendrick (1927).
\newblock A contribution to the mathematical theory of epidemics.
\newblock {\em Proceedings of the Royal Society of London. Series A\/}~{\em
  115\/}(772), 700--721.

\bibitem[\protect\citeauthoryear{King, Domenech~de Cell{\`e}s, Magpantay, and
  Rohani}{King et~al.}{2015}]{king2015avoidable}
King, A.~A., M.~Domenech~de Cell{\`e}s, F.~M. Magpantay, and P.~Rohani (2015).
\newblock Avoidable errors in the modelling of outbreaks of emerging pathogens,
  with special reference to {Ebola}.
\newblock {\em Proceedings of the Royal Society B: Biological Sciences\/}~{\em
  282\/}(1806), 20150347.

\bibitem[\protect\citeauthoryear{Lin, Lin, Chiu, and He}{Lin
  et~al.}{2016}]{lin2016seasonality}
Lin, Q., Z.~Lin, A.~P. Chiu, and D.~He (2016).
\newblock {Seasonality of influenza A (H7N9) virus in China-—fitting simple
  epidemic models to human cases}.
\newblock {\em PLoS one\/}~{\em 11\/}(3), e0151333.

\bibitem[\protect\citeauthoryear{Martinez, King, Yunus, Faruque, and
  Pascual}{Martinez et~al.}{2016}]{martinez2016differential}
Martinez, P.~P., A.~A. King, M.~Yunus, A.~Faruque, and M.~Pascual (2016).
\newblock Differential and enhanced response to climate forcing in diarrheal
  disease due to rotavirus across a megacity of the developing world.
\newblock {\em Proceedings of the National Academy of Sciences\/}~{\em
  113\/}(15), 4092--4097.

\bibitem[\protect\citeauthoryear{Pons-Salort and Grassly}{Pons-Salort and
  Grassly}{2018}]{pons2018serotype}
Pons-Salort, M. and N.~C. Grassly (2018).
\newblock Serotype-specific immunity explains the incidence of diseases caused
  by human enteroviruses.
\newblock {\em Science\/}~{\em 361\/}(6404), 800--803.

\bibitem[\protect\citeauthoryear{Schenzle}{Schenzle}{1984}]{schenzle1984age}
Schenzle, D. (1984).
\newblock An age-structured model of pre-and post-vaccination measles
  transmission.
\newblock {\em Mathematical Medicine and Biology: A Journal of the IMA\/}~{\em
  1\/}(2), 169--191.

\bibitem[\protect\citeauthoryear{Ster, Singh, and Ferguson}{Ster
  et~al.}{2009}]{ster2009epidemiological}
Ster, I.~C., B.~K. Singh, and N.~M. Ferguson (2009).
\newblock {Epidemiological inference for partially observed epidemics: the
  example of the 2001 foot and mouth epidemic in Great Britain}.
\newblock {\em Epidemics\/}~{\em 1\/}(1), 21--34.

\bibitem[\protect\citeauthoryear{Webb, Browne, Huo, Seydi, Seydi, and
  Magal}{Webb et~al.}{2015}]{webb2015model}
Webb, G., C.~Browne, X.~Huo, O.~Seydi, M.~Seydi, and P.~Magal (2015).
\newblock {A model of the 2014 Ebola epidemic in West Africa with contact
  tracing}.
\newblock {\em PLoS currents\/}~{\em 7}.

\bibitem[\protect\citeauthoryear{Xia, Bj{\o}rnstad, and Grenfell}{Xia
  et~al.}{2004}]{xia2004measles}
Xia, Y., O.~N. Bj{\o}rnstad, and B.~T. Grenfell (2004).
\newblock Measles metapopulation dynamics: a gravity model for epidemiological
  coupling and dynamics.
\newblock {\em The American Naturalist\/}~{\em 164\/}(2), 267--281.

\bibitem[\protect\citeauthoryear{Yang, Chen, and Zhang}{Yang
  et~al.}{2013}]{yang2013stability}
Yang, J.-Y., Y.~Chen, and F.-Q. Zhang (2013).
\newblock Stability analysis and optimal control of a hand-foot-mouth disease
  ({HFMD}) model.
\newblock {\em Journal of Applied Mathematics and Computing\/}~{\em 41\/}(1-2),
  99--117.

\end{thebibliography}
\end{document}